\documentstyle[pra,aps,floats,psfig]{revtex}

\begin{document}

\draft

\title{Nonadiabatic dynamics in evaporative cooling of trapped atoms
by a radio frequency field}

\author{K.-A. Suominen}
\address{Helsinki Institute of Physics, PL 9, FIN-00014 Helsingin
yliopisto, Finland}
\address{Theoretical Physics Division, Department of Physics,
University of Helsinki, PL 9, FIN-00014 Helsingin yliopisto, Finland}

\author{E. Tiesinga\cite{Maryland} and P. S. Julienne}
\address{Atomic Physics Division,
National Institute of Standards and Technology, Gaithersburg,
MD 20899-0001}

\date{\today}

\maketitle

\begin{abstract}
Magnetically trapped neutral atoms can be cooled with the evaporation
technique.  This is typically done by using a radiofrequency (rf) field
that adiabatically couples trapped and untrapped internal atomic states
for atoms with kinetic energies above a value set by the field frequency.
The rf-field can also induce nonadiabatic changes of internal atomic spin
states ($F,M$) that lead to heating and enhanced loss of atoms.  In this
paper we use wave packet simulations to show that the evaporation process
can induce these nonadiabatic transitions which change the internal spin
state of doubly spin-polarized (2,2) trapped atoms.  We also verify
the validity of a multistate Landau-Zener model in describing the
nonadiabatic dynamics.  In addition, we calculate exchange relaxation
rate coefficients for collisions between atoms in the ($2,M$) states of
$^{23}$Na atoms.  Large exchange relaxation coefficients for $^{23}$Na
as compared to $^{87}$Rb $F=2$ suggest that evaporative cooling of (2,2)
Na will be more difficult than for the corresponding state of~Rb.
\end{abstract}
\pacs{34.90.+q, 32.80.Pj, 42.50.Vk}

\section{Introduction}

Bose-Einstein condensation has been observed recently in alkali
gases~\cite{Anderson95,Davis95,Mewes96a,Bradley97}. This has led
to a series of fascinating experiments involving collective
excitations~\cite{Jin96,Edwards96,Mewes96b,Jin97,Andrews97a} and atom
lasers~\cite{Mewes97,Andrews97b}. These experiments take place at very
low temperatures, ranging from microkelvins to nanokelvins. Such temperatures
can be reached only with evaporative cooling if sufficient atomic densities
for condensation are to be obtained. The idea for evaporative cooling was
originally developed in connection with attempts to observe Bose-Einstein
condensation in spin-polarized hydrogen~\cite{Hess86,Tommila86} (for a review
see Ref.~\cite{Ketterle96}). Although the process has been studied in detail
both theoretically and experimentally, some aspects still require
clarification.

\begin{figure}[thb]
\centerline{
\psfig{width=120mm,file=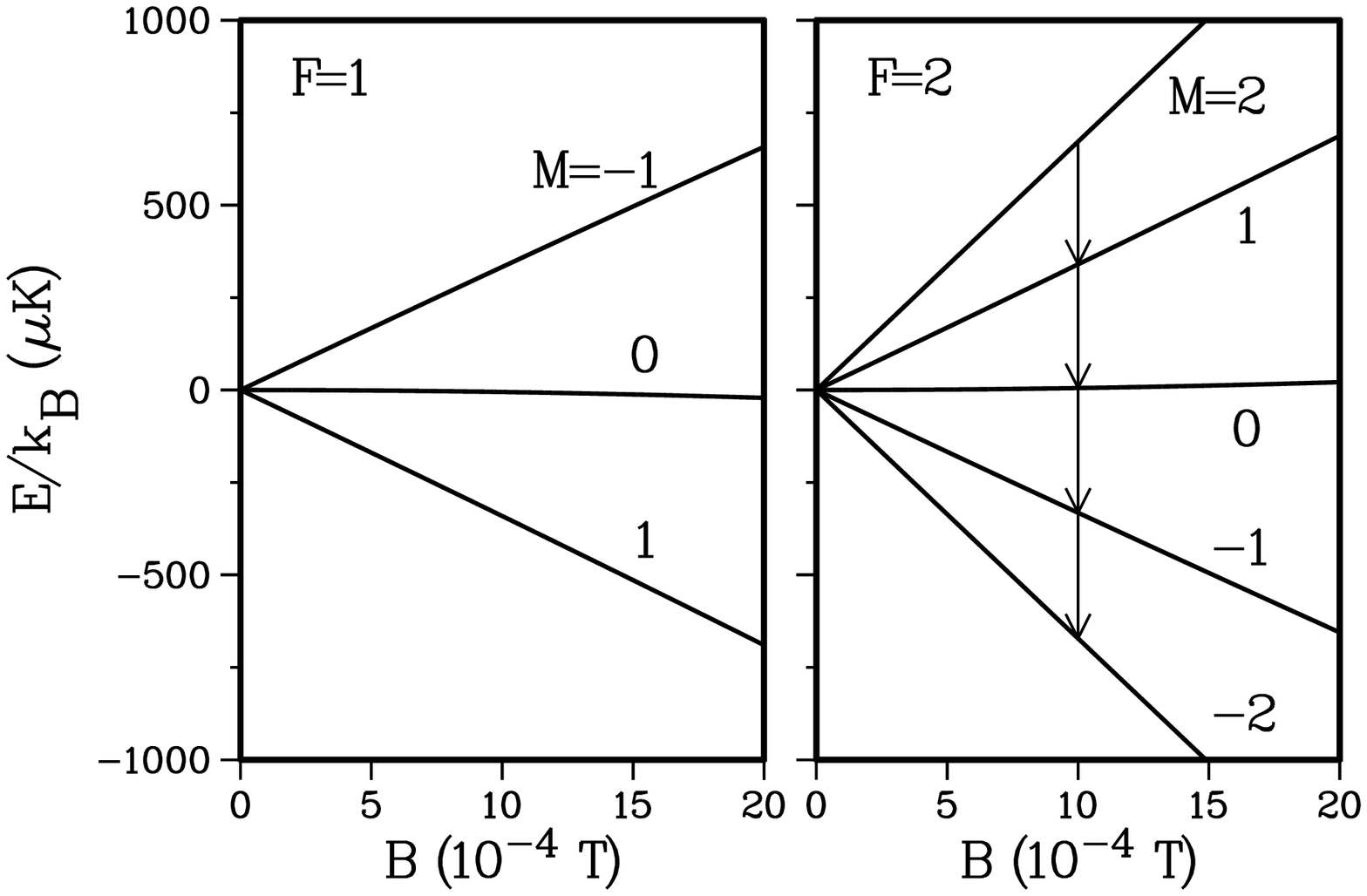}
}
\vspace{2mm}
\caption[f0]{The hyperfine structure of the ground state for $^{23}$Na
as a function of magnetic field $B$.  In a typical trapping situation
the $F=1$, $M=-1$ or $F=2$, $M=2$ state is used, $B$
depends quadratically on a spatial coordinate and is on the order of
1~mT in the middle of the trap.  The left frame shows
the magnetic field dependence of the three $F=1$ magnetic sublevels.
For fields up to 2~mT the internal energy is in the linear Zeeman
regime.  The right frame shows the five $F=2$ sublevels.  The arrows
indicate the rf-induced coupling between the various sublevels for an
atom initially in the $F=2$, $M=2$ state, where
the length of the arrow is proportional to the rf frequency $\nu$.
\label{idea}}
\end{figure}

\begin{figure}[thb]
\centerline{ \psfig{width=86mm,file=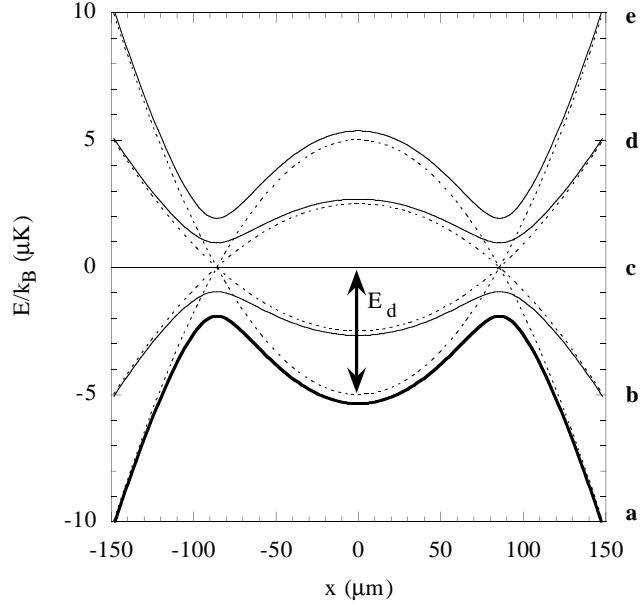} } \vspace{2mm} 
\caption[f1]{Bare potentials (dotted lines) and corresponding 
rf-dressed potentials (solid lines) for the $^{23}$Na $F=2$ system.  
After performing the rotating wave approximation the 
atomic states are shifted by multiples of the rf-field quantum and 
rf-resonances occur as crossings.  The trap frequency is 112 Hz.  The 
bare state trap depth is $E_d/k_B=5\ \mu$K. The rf-field is resonant 
with the $M$-states at $x\simeq 86\ \mu$m.  The bare potentials are at 
the center of the trap $M=2,1,0,-1,-2$, in the order of increasing 
energy.  Correspondingly the rf-dressed states are labelled from $a$ 
to $e$. The trapping state $a$ is indicated by a thick solid line.
\label{pots}}
\end{figure}

The basic idea for evaporative cooling is simple.  The atoms in a
hyperfine state ($F,M$) are trapped by spatially inhomogeneous
magnetic fields.  The edges of the trap are then lowered, allowing the
hot atoms to escape.  Atoms left in the trap collide elastically and
eventually a thermal equilibrium is reached, but at some temperature
which is lower than the original one.  In practice the trap edges are
"lowered" by coupling different magnetic substates (spin states) of
the atoms with a radiofrequency (rf) field of variable
frequency~\cite{Ketterle96,Pritchard88}, as shown in Fig.~\ref{idea}.
Within the rotating wave approximation we can shift the energy states
by multiples of the radiofrequency field photon energy, and eventually
we obtain the curve crossing situation for $F=2$ featured in
Fig.~\ref{pots}~\cite{Ketterle96}.  We can view the evaporation as
adiabatic following of the lowest rf-dressed state by the moving
atoms.  Here we will consider the doubly spin-polarized level
($2,2$), for which Bose-Einstein condensation has been achieved for
$^{87}$Rb but not for $^{23}$Na.

The evaporation process requires elastic binary collisions for the
rethermalisation of the atoms that remain trapped.  However, inelastic
collisions can give the collision products large kinetic energies
which lead to loss or heating of trapped atoms.  The normal mechanism
for collisional relaxation of the trapped (2,2) doubly spin-polarized
state of $^{87}$Rb or $^{23}$Na, spin-dipolar relaxation, has a very
small collisional rate coefficient~\cite{Stoof89,Tiesinga91,Mies96}
which would limit the lifetime of a Bose-Einstein condensate to a very
acceptable 30 seconds.  However, if the evaporation process results in
the formation of ($2,1$) and ($2,0$) spin components via rf-induced
nonadiabatic transitions due to atomic motion in the spatially
inhomogeneous magnetic field, inelastic collisions between these
components and/or the remaining (2,2) spin state can occur by a strong
spin-exchange mechanism and in principle lead to large collisional
relaxation rates.

The actual size of the relaxation rate is strongly species dependent. In
Refs.~\cite{Kokkelmans97,Julienne97} it has been shown that spin relaxation
for the collision between (2,2) and ($1,-1$) for $^{87}$Rb is strongly
reduced compared with the equivalent collision in $^{23}$Na.
The rate coefficients are anomalously small for $^{87}$Rb,
because of the fortuitous near equality of the scattering lengths for the
$^3\Sigma_u^+$ and $^1\Sigma_g^+$ states that correlate with the
separated ground state atoms.  We will also find in this paper that the
various spin-exchange collisional rate coefficients between $^{23}$Na
$(2,M)$ spin components are about three orders of magnitude larger than
the corresponding values for $^{87}$Rb.

In this paper we consider two distinct aspects of the relevant
physics: (1) the rf-induced nonadiabatic dynamics due to motion in the
spatially inhomogeneous magnetic field that move atoms from the
primary trapped state (2,2) to the other states of the same hyperfine
manifold, and (2) the rate coefficients for collisions involving other
($2,M$) states.  For the first aspect, we consider the coupled
five-state multicrossing situation that occurs in the rf-dressed
picture for $^{23}$Na ($2,M$) shown in Fig.~\ref{pots}.  Especially in
the early stages of cooling, when the frequency of the rf-field is
large and the crossing between the ($2,M$) states occurs at very large
energies, it is possible that the atoms will not adiabatically follow
the rf-dressed potential $a$ shown in Fig.~\ref{pots}.  Atoms can,
after traversing the crossing, either remain on the bare (2,2)
state~\cite{bare} or move, for example, to the bare (2,1) state.  In
other words, we have nonadiabatic dynamical transfer between the
rf-dressed states.  These two bare states are trapping states, so
atoms will return to the crossing region instead of escaping from the
trap.  When they traverse the crossing again, population can move to
the spin states (2,1) and (2,0), and then approach the trap center,
thus leading to the loss-inducing collisions to be considered as the
second aspect of the physics.

The adiabatic following, leading to evaporation, and the unwanted nonadiabatic
transitions can be studied by describing atoms with multicomponent wave
packets that evolve under a time-dependent Schr{\"o}dinger equation.
This model does not contain all aspects of atomic dynamics during evaporation
such as the rethermalization, but
does demonstrate adequately the possibility of nonadiabatic transitions
and the prospects for harmful inelastic atomic collisions in the
trap.  It has been proposed that a Landau-Zener model can be used
for a semiclassical study of the evaporation as well~\cite{Ketterle96}
and we will show that a multistate version of the Landau-Zener
model~\cite{Kazansky96,Ostrovsky97,Vitanov97} describes our wave packet
dynamics rather well.  This allows one to extrapolate some of our results into
a much larger parameter space than we have covered with our wave packet
analysis.

The lowest rf-dressed state has nonnegligible contributions from all ($2,M$)
bare states of the $F=2$ manifold. This is called rf-mixing of the spin states.
One might think that the collisions between these mixed states can effectively
lead to collisions between the different $M$-states, especially when the energy
difference between the states becomes nonnegligible compared to the strength
of the rf-coupling.  However, we find that this mixing will not lead to
enhanced inelastic collision rates at all. This conclusion agrees with that
obtained by Moerdijk {\it et al.}~\cite{Moerdijk96}.  The physical reason for
this behavior is that the rf-field only rotates the quantization axis, and
the $T$-matrix elements for the inelastic processes change accordingly, not
picking up any exchange collision amplitude.  Thus the rf-dressed state which
has the lowest energy simply obtains the label (2,2) along the new
quantization axis.  In short, when two atoms in the lowest rf-dressed state
collide, the process is equal to a collision between two atoms in the bare
(2,2) state.

In this paper we first present the collisional rate coefficients for $^{23}$Na 
$F=2$ trapping in Sec.~\ref{rates} and discuss the rf-mixing of states. Then 
we demonstrate the nonadiabatic dynamical transfer processes in
Sec.~\ref{nonadiab} using wave packets.  Furthermore, we discuss the 
multistate Landau-Zener modelling of the dynamics of the evaporation process.  
Finally, the discussion in Sec.~\ref{conclusions} concludes the paper.

\section{Collisional loss rates}\label{rates}

\subsection{Exchange relaxation rates}\label{exchange}

The $^{23}$Na and $^{87}$Rb ground state atoms have two hyperfine
components, $F=1$ and $F=2$ as shown for $^{23}$Na in Fig.~\ref{idea}.
The $F=1$ level is 3-fold degenerate, with components $M = 1,0,-1$ and
the $F=2$ level is 5-fold degenerate, with components $M=-2,-1,0,1,2$.
For magnetic fields below $\sim$ 1 mT the internal energy of the magnetic
sublevels depends linearly on the trapping field where the magnetic sublevels
are defined with respect to the direction of this field.
The $F$ label is strictly speaking only valid for zero magnetic field.
However, we wish to determine the inelastic collision rate coefficients in a
weak (mT) magnetic field and it turns out that the dominant loss
processes are nearly independent of field strength $B$ and are well
approximated by the corresponding zero-field rate. We will present $B=0$ rates
and scattering $T$-matrix elements labeled according to the zero-field labels
($F,M$) only, but have confirmed numerically that small magnetic fields do
not change the rate by more than 20~\%.

The (2,2) trapping state is known to
relax with a small rate coefficient, on the order of $10^{-15}$ cm$^3$/s,
due to the weak spin-dipole mechanism~\cite{Tiesinga91,Mies96,Stoof88}.
Relaxation leads to formation of a lower $(1,M)$ state accompanied by
the release of a large amount of energy.
However, if other ($2,M$) components of $F=2$ are present due to
rf-induced nonadiabatic motion in the trapping potential, these
components may decay via the exchange mechanism.  We will show that
this mechanism leads to very large rate coefficients for $^{23}$Na, 
greater than $10^{-11}$ cm$^3$/s, that is, orders of magnitude larger
than for dipolar relaxation.  On the other hand, ($2,M$) components of
$^{87}$Rb have very small exchange relaxation rate coefficients, on
the order of $10^{-14}$ cm$^3$/s, just as is known to be the case for
exchange collisions between a ($2,2$) and ($1,-1$)
atom~\cite{Kokkelmans97,Julienne97}.  The small rate coefficients for
$^{87}$Rb ensure that this process is not significant for the $F=2$
level of $^{87}$Rb, which can be evaporatively cooled and
Bose-condensed in the ($2,2$) state~\cite{Myatt97}.  So far no one
has achieved Bose-Einstein condensation for the ($2,2$) level of $^{23}$Na.

We now extend the calculations of Ref.~\cite{Julienne97} to apply to
collisions of two ($2,M$) atoms as well.  We assume that the collision 
starts with the two atoms in the
hyperfine levels ($F_1,M_1$), ($F_2,M_2$) and with relative angular
momentum state $\ell,m$ and ends with the atoms in the levels
($F_3,M_3$), ($F_4,M_4$) and relative angular momentum state
$\ell',m'$. At sufficiently low temperature only $s$-waves
contribute, for which $\ell=0$, $m=0$.  We introduce symmetrized
states of atom pairs that are symmetric with respect to exchange of
identical atoms, with half-integer nuclear spin quantum number, and
define a symmetrized $T$-matrix as done by Stoof {\it et al.}~\cite{Stoof88}
for transitions between the symmetrized states
\begin{equation}
   |\{F_iM_i,F_jM_j\}\ell m\rangle =
   \frac{|F_iM_i,F_jM_j,\ell m\rangle + (-1)^\ell
   |F_jM_j,F_iM_i,\ell m\rangle} {\sqrt{2(1+\delta_{ij})}},
   \label{rf.2b}
\end{equation}
where $\delta_{ij}=$ 1 if $F_i=F_j,M_i=M_j$ and $\delta_{ij}=0$ otherwise.

Alternatively this basis is rewritten by noting that the total angular
momentum,
$\vec{F}_t=\vec{F}_1+\vec{F}_2+\vec{l}= \vec{f}+\vec{l}$, is conserved
for $B=0$.  Therefore, after vector coupling the $|F_1M_1\rangle$ and
$|F_2M_2\rangle$ levels to the resultant symmetrized angular
momentum $|F_1F_2f,M_1+M_2\rangle$ states, then vector coupling to
the relative angular momentum $|\ell,m\rangle$ to get the total
angular momentum states $|F_1F_2f,\ell,F_t, M_t=M_1+M_2+m
\rangle$, the symmetrized $T$-matrix elements are given in
terms of the indices $\{F_1F_2f \ell F_t\}$ and are independent of
projection quantum numbers $M_1,M_2, m, M_t$.  The $T$-matrix
elements for the symmetrized states of Eq.~(\ref{rf.2b}) are given in
terms of these $T$-matrix elements in the total angular momentum basis
as~\cite{Mies73,Baylis87,Williams}
\begin{eqnarray}
   && \langle \{F_3M_3,F_4M_4\}\ell'm' | T |
      \{F_1M_{1},F_2M_2\}\ell m\rangle =\nonumber\\
   &&
   \left(\frac{1+\delta_{F_1F_2}}{1+\delta_{F_1F_2}\delta_{M_1M_2}}
       \right)^{\frac{1}{2}}
   \left(\frac{1+\delta_{F_3F_4}}{1+\delta_{F_3F_4}\delta_{M_3M_4}}
       \right)^{\frac{1}{2}}
       \sum _{ff'F_t}
      (F_3F_4f'|M_3M_4,M_3+M_4)(f'\ell' F_t|M_3+M_4,m',M_t) \nonumber\\
   && \times (F_1F_2f|M_1M_2,M_1+M_2)(f\ell F_t|M_1+M_2,m,M_t)
      T^{(F_t)}(F_3F_4f'\ell'\leftarrow F_1F_2f\ell) ,
      \label{rf.4a}
\end{eqnarray}
where $(\ |\ )$ are the Clebsch-Gordan coefficients.  We define our
$T$-matrix to be related to the unitary $S$-matrix by $\bf T=1-S$.

Spin-exchange collisions are only possible if the following selection
rules are obeyed: $\ell=\ell'$, $m=m'$, $f=f'$, and $M_1+M_2=M_3+M_4$.
Equation~(\ref{rf.4a}) simplifies considerably for $s$-wave collisions,
since $\ell=0$ and $f'=f=F_t$; in particular, for $s$-wave $(2,M_1)+(2,M_2)$
collisions inelastic exchange transitions are only possible if $f=F_t= 0$
or 2.

The spin-exchange selection rules show that a doubly-polarized gas cannot 
decay via this mechanism. However, if the
evaporation process produces the other trappable $F=2$ states this is no
longer valid. Table~\ref{table1} therefore gives explicit formulas that can
be used for evaluating the $T$-matrix amplitudes for $(2,2)+(2,0)$ and
$(2,1)+(2,1)$ $s$-wave exchange collisions.  The shorthand notation used in
Table~\ref{table1} is
\begin{eqnarray}
   T_{21}^{(2)} &=& T^{(2)}(212s \leftarrow 222s),
   \label{rf.5a} \\
   T_{11}^{(2)} &=& T^{(2)}(112s \leftarrow 222s), \label{rf.5b}
\end{eqnarray}
where the notation means $T_{F_3F_4}^{(F_t)}$.  Amplitudes for other
$(2,M_1)+(2,M_2)$ $s$-wave collisions can be worked out using
Eq.~(\ref{rf.4a}) and the additional matrix element,
\begin{equation}
   T_{11}^{(0)} = T^{(0)}(110s \leftarrow 220s). \label{rf.5c}
\end{equation}

One might expect that $(2,2)+(2,1)$ inelastic exchange collisions will be
important.  However, since then $M_3+M_4=M_1+M_2=3$, it follows that
spin-exchange matrix elements with $f=f'=3$ or 4 contribute.
For $s$-wave scattering there is only one spin state with either value of $f$.
Hence only elastic $s$-wave $T$-matrix elements exist while
inelastic $s$-wave spin-exchange transitions are absent.
The only remaining $(2,2)+(2,1)$ relaxation mechanism is via the weak dipolar
interaction.
On the other hand a $p$-wave exchange relaxation is permitted for $(2,2)+(2,1)$
collisions, where the inelastic $T^{(F_t)}(213p\leftarrow 223p)$ matrix
elements
contribute as the temperature increases.  These $T$-matrix elements vary
as $k^{3/2}$, where $\hbar k$ is the relative momentum, whereas the
$s$-wave amplitudes vary as $k^{1/2}$.  In numerical calculations, we
find $p$-wave inelastic exchange collisions will have rate coefficients two
orders of magnitude or more smaller than $s$-wave exchange collisions
for temperatures below 10 $\mu$K for $^{23}$Na collisions.  The $p$-wave
contributions rapidly increase with collision energy, and become
comparable to the $s$-wave contributions above 100 $\mu$K. Thus, some
collisional relaxation due to $(2,2)+(2,1)$ collisions may be
important at the start of evaporation, but such collisions become less
significant as the condensation limit is approached.

\begin{table}[tbp]
\begin{tabular}{rr}
   $\{F_3M_3,F_4M_4\}\leftarrow\{2M_1,2M_2\}$ & $\langle \{F_3M_3,F_4M_4\} 
   s|T| \{2M_1,2M_2\}s\rangle$ \\  \hline
   $21,11\leftarrow 22,20$ & $-\sqrt{2} \sqrt{\frac{2}{21}} T_{21}^{(2)}$ \\
   $11,11\leftarrow 22,20$ & $\sqrt{2} \sqrt{\frac{2}{7}} T_{11}^{(2)}$ \\
   $22,10\leftarrow 22,20$ & $\sqrt{2} \frac{2}{\sqrt{21}} T_{21}^{(2)}$ \\
   $21,11\leftarrow 21,21$ & $\sqrt{\frac{1}{7}} T_{21}^{(2)}$ \\
   $11,11\leftarrow 21,21$ & $-\sqrt{\frac{3}{7}} T_{11}^{(2)}$ \\
   $22,10\leftarrow 21,21$ & $-\sqrt{\frac{2}{7}} T_{21}^{(2)}$ \\
\end{tabular}
\vskip0.5cm
\caption{Bare exchange scattering $T$-matrix elements for $s$-waves}
\label{table1}
\end{table}

In order to calculate the actual zero-field exchange collision rates for
collisions of $^{23}$Na ($2,M$) levels, we use the same quantum
scattering method and model of the $^{23}$Na potentials as in
Ref.~\cite{Tiesinga96}.  We adjusted the potentials slightly to locate
correctly the positions of the triplet Feshbach resonance levels
reported recently~\cite{Inouye98}; the calculated scattering length
for $(1,-1)$ elastic collisions is 52 a$_0$, in agreement with our
previously reported value~\cite{Tiesinga96}.  We find as $k\rightarrow
0$,
\begin{eqnarray}
   T_{21}^{(2)}&=&5.5e^{-3.1i}\sqrt{k},\label{rf.6a}\\
   T_{11}^{(2)}&=&3.9e^{-1.4i}\sqrt{k}, \\
   T_{11}^{(0)}&=&6.5e^{-1.6i}\sqrt{k} ,\label{rf.6b}
\end{eqnarray}
where $k$ is expressed in units of a$_0^{-1}$ (a$_0$ = 0.0529177 nm is
the Bohr radius of the hydrogen atom).  We define $|T_{ij}^{(f)}|^2 =
A^{(f)}_{ij}k$ and have $A^{(2)}_{21}=$ 31 a$_0$, $A^{(2)}_{11}=$ 15
a$_0$, and $A_{11}^{(0)}$ = 43 a$_0$.
On the other hand, the squared $T$-matrix for inelastic spin-exchange $p$-wave
scattering is proportional to $k^3$. In particular,
$|T^{(3)}(213p\leftarrow 223p)|^2= A^{(3)}_{21} k= (2.6\times10^5 k^2)k$.
This corresponds to $A^{(3)}_{21}=$ 0.3 a$_0$ at 10 $\mu$K which as
noted is small compared to inelastic $s$-wave contributions at this collision
energy.

The event rate coefficient due to any atom loss process at a
relative collision momentum $\hbar k$, using the definition $|T|^2=Ak$, is
conveniently written as
\begin{equation}
   K_{\rm event} = \frac{\pi v}{k^2} |T|^2 = \frac{\pi \hbar}{\mu} A
   = 1.056 \times 10^{-11} \frac{A({\rm a}_0)}{\mu({\rm amu})} \,\,
   \frac{\rm cm^3}{\rm s}
   \label{rf.7}
\end{equation}
for a Maxwellian gas and half of this for a Bose-Einstein
condensate~\cite{Stoof89}.  The quantity $\mu$ is the reduced mass and the
number of hot atoms produced per event in
either case is two, since the released energy is shared equally
between the two atoms.

Using Eqs.~(\ref{rf.6a}), (\ref{rf.6b}) and Table~\ref{table1}, the $A$
coefficients in Eq.~(\ref{rf.7}) for the total event rate of spin-exchange
collisions for $(2,2)+(2,0)$ and $(2,1)+(2,1)$ are
$\frac{3}{7}(A^{(2)}_{21}+A^{(2)}_{11})=$ 20 a$_{0}$ and $\frac{2}{7}
(2A^{(2)}_{21}+A^{(2)}_{11})=$ 22 a$_{0}$, respectively.  The event
rate coefficients for these collisions are on the order of $2 \times 10^{-11}$
cm$^{3}$/s for Na.  The rate coefficients are in
excellent agreement with those calculated numerically for weak
magnetic fields in the range 0~mT to 20~mT.

The corresponding inelastic collision strengths $A$ for $^{87}$Rb are
$A^{(2)}_{21}=$ 0.064
a$_0$, $A^{(2)}_{11}=$ 0.37 a$_0$, and $A_{11}^{(0)}$ = 0.16 a$_0$ for
$s$-waves and $A^{(3)}_{21}=$ $3.7 \times 10^4 k^2$ for $p$-waves.
These values are obtained as in Ref.~\cite{Julienne97} with ground
state potentials that give scattering lengths of 103 a$_0$ for both the
$(2,2) + (2,2)$ and $(1,-1) + (1,-1)$ collision.
The $s$-wave event rate coefficients are on the
order of a few times $10^{-14}$ cm$^3$/s, consistent with the small
exchange collision rate coefficient of comparable magnitude for
$^{87}$Rb $(2,2) + (1,-1)$
collisions~\cite{Kokkelmans97,Julienne97,Myatt97,Burke97}.  The $^{87}$Rb
exchange collision rate coefficients show stronger sensitivity to
weak magnetic field than the $^{23}$Na ones\cite{Julienne97}.

For Bose-Einstein
condensation studies the typical atomic densities are on the order of
$10^{14}$ atoms/cm$^{3}$.
Therefore the three orders of magnitude difference between $^{23}$Na and
$^{87}$Rb rate coefficients implies that
even relatively small population transfer from the (2,2) trapped level
to the $(2,1)$ and $(2,0)$ states during the evaporation process
can cause a strong/unwanted increase in trap loss for
$^{23}$Na, but has an insignificant effect for $^{87}$Rb.

\subsection{Effect of the rf-field on exchange relaxation}

The rate coefficients in Sec.~\ref{exchange} are evaluated using
so-called bare atomic states where the rf-field has not been included.
We now turn our attention to the effect of rf-induced dressing on the
inelastic collision rate coefficients.  The dressing mixes
different ($2,M$) levels and there is a question as to whether this
mixing affects the exchange relaxation rates of the dominant (2,2)
trapped level.  Moerdijk {\it et al.}~\cite{Moerdijk96} have argued
that field-dressing has a negligible effect on collisional relaxation.
We now discuss the rf-mixing and its consequences.

In a magnetic trap the atoms are held in a position dependent trapping field
$\vec{B}_{\rm trap}$. In fact, for a spherically harmonic trapping potential
$|\vec{B}_{\rm trap}(\vec{x}) - \vec{B}_{\rm trap}(\vec{0})|\propto x^2$,
where $\vec{x}=\vec{0}$ is the center of the trap.  This magnetic field
defines a space-fixed $z$-axis at each point in the trap along which
the atomic ($F,M$) states are defined. The variations of the magnetic
field strength over the atom cloud are small compared to the field strength
in the middle of the trap.
Hence if only this trapping field is present the zero-field $T$-matrix
elements and the corresponding rate coefficients are valid everywhere in the
trap.

The components of an rf-field, with frequency $\nu$, parallel and
perpendicular to the trapping field have different effects on the mixing
of states~\cite{Vanier89}.
The latter leads to $|\Delta M|=1$ transitions. In fact, for the
evaporation process the rf-photon energy $h\nu$ is chosen to be nearly
resonant with the ($2,2$) to ($2,1$) transition and since the trapping
field is sufficiently weak that the internal energy of the atomic states
depends linearly on the field strength the photon is resonant with all
adjacent $M$ states as well.  The parallel component of the rf-field
leads to $|\Delta M|=0$ transitions. However, the rf-photon
is not resonant with the allowed parallel transitions ($1,M$) $\to$ ($2,M$).
Therefore, at each point in the trap only the component of the rf-field
that is locally perpendicular to the trapping field leads to mixing
of atomic sublevels. The amplitude of the rf-field is small compared to the
trapping field.

In order to keep the modelling tractable we make several although not
crucial simplifications.
Variations in the direction of the trapping field over the atom cloud can be
neglected. This allows us to use a single $z$-axis that describes the
quantization axis everywhere in the trap.
Moreover, the atom-field interaction will be treated using the rotating wave
approximation. Then it is
convenient to use a coordinate system that rotates with frequency $\nu$
around the $z$-axis. The rf-field is stationary in this coordinate system
and is chosen to lie along the rotating $x'$-axis with a field strength
$B_{\rm rf}$. Following Ref.~\cite{Vanier89} the $F=2$ atomic hyperfine
Hamiltonian coupling to the rf-field can then be interpreted in terms of a
magnetic moment $\vec{F}$ rotating around an effective field and
we arrive at a local Hamiltonian which describes the
motion of the magnetic moment at each position $\vec{x}$ in the magnetic trap
\begin{equation}
   H(\vec{x}) = \hbar\sigma(\vec{x}) \hat{F}_z +\hbar\Omega
                \hat{F}_{x'},\label{sigm2}
\end{equation}
where $\hbar\sigma$ is the local energy spacing of the atomic magnetic
sublevels, shown as the dotted lines of Fig.~\ref{pots}, and $\Omega$ is the
zero-detuning Rabi frequency of the atom-field coupling.  Here
$\hat{F}_z$ and $\hat{F}_{x'}$ are the components of spin-$F$ operators
along the $z$-~and $x'$-axis respectively.
Note that for example in the space-fixed $x$ direction it follows
\begin{equation}
   \hbar\sigma=\frac{1}{F}
   \left(\frac{1}{2}m\omega^2x^2-E_d\right),\label{sigma}
\end{equation}
with $x=0$ as the trap center, $m$ is the atomic mass and $\omega$ is the
trap frequency in the $x$ direction.  Here $E_d$ is the bare state trap
depth (see Fig.~\ref{pots}), defined as the difference between the bare level
spacing of adjacent $M$ states in the center of the trap and the rf frequency
$h\nu$.

We have assumed for simplicity that the trapping in three dimensions is
achieved by a static field~\cite{Davis95}, and not a TOP
trap~\cite{Anderson95}.
In a TOP trap there is an additional
fast-oscillating field which has an amplitude comparable to the trapping field.
The amplitude of the rf-field used for evaporation is small compared to the
trapping field (even at the center of the trap where the trapping field is
weakest), which allows the description we presented above. If we
simply assume that the fast-oscillating TOP field can be adiabatically
eliminated, i.e., we obtain the time-averaged trapping potential, our studies
apply also to the TOP trap. The validity of this adiabatic elimination of the
TOP field in the presence of the evaporation rf-field is to some extent an open
question~\cite{BillKris}. It does not fall within the scope of this paper,
however, so we do not discuss it further.

A collision between two atoms occurs locally in sub-nanosecond time scales.
It is sufficiently fast that the rf-field as seen by the atoms during a
collision is frozen~\cite{Moerdijk96} (the period of the rf-field is on the
order of about 100 ns). Thus the rotating $x'$-axis can be considered
fixed during the collision.  Therefore, Eqs.~(\ref{rf.4a})-(\ref{rf.7}) are
again valid but now with spin states $|FM\rangle$ defined relative to a
rotating $z'$-axis. To see this just note that the simple
spin structure of Eq.~(\ref{sigm2}) allows us to introduce a transformation
$R$ that rotates the $z$-axis into a $z'$-axis with the property that
the transformed Hamiltonian is of the form $\Delta(\vec{x}) F_{z'}$. In fact,
it can be shown that this is achieved with a rotation by an angle $\theta$
around the axis perpendicular to both $z$- and $x'$-axis, where
\begin{equation}
   \tan(\theta) = -\frac{\Omega}{\sigma}
\end{equation}
and $\Delta(\vec{x})=\hbar\sqrt{\sigma(\vec{x})^2+\Omega^2}$.
The explicit form of the rotation is
\begin{equation}
    | FM \rangle_{z'} = \sum_{M'} d^{(F)}_{MM'}(-\theta)| FM' \rangle_z,
\end{equation}
where the $d^{(F)}_{MM'}$ are rotation matrices~\cite{Zare88} and the 
subscripts $z$ and $z'$ denote the axis along which the state 
$|FM\rangle$ is defined.  Again this $z'$-axis can be considered fixed 
during a collision.  Consequently, if the atoms before the collision 
are in an eigenstate of the Hamiltonian~(\ref{sigm2}), the decay 
processes due to collisions are simply given with the bare rate 
constants.  For example, if the atoms are everywhere in the local 
doubly-polarized state $| FM \rangle_{z'}$, only the weak spin dipole 
interaction causes decay of the sample. We have explicitly verified e.g. 
for the dressed (2,2) state that the exchange transition contributions 
from the bare states such as (2,1) and (2,0) cancel out completely.

The premise that the atoms are always in the eigenstates of the 
Hamiltonian~(\ref{sigm2}) hinges on the assumption that the motion of 
the atoms has a negligible effect.  In fact the premise is not always 
satisfied and Sec.~\ref{nonadiab} is devoted to the nonadiabatic 
effects induced by the motion of a single atom through a spatially 
varying magnetic field and hence a varying $\sigma$.  In such a field 
at each point in the trap the coordinate system with the quantization 
axis $z'$ is different.  This is most inconvenient if we want to 
follow the motion of a single atom.  Therefore, we use the basis of 
eigenstates $|FM\rangle_z$ of the Hamiltonian~(\ref{sigm2}) when 
$\Omega=0$ everywhere in the trap; these are the base states~\cite{bare}.  
Of course, for non-zero $\Omega$ 
the Hamiltonian is not diagonal in this basis.  There is coupling of 
adjacent magnetic levels with
\begin{equation}
   H_{M,M+1}=H_{M+1,M}=\sqrt{(F-M)(F+M+1)}\hbar\Omega/2.\label{coupl}
\end{equation}

\section{Dynamics of evaporation and nonadiabatic transitions}\label{nonadiab}

The calculation of the transition amplitudes in Sec.~\ref{rates} shows that for
$^{23}$Na the collisions between the atoms in an rf-field does not
necessarily lead to enhanced inelastic rate coefficients. However, if we
consider moving atoms in the absence of collisions, then for an atom initially
in the (dressed) doubly polarized spin state it is possible to have
nonadiabatic transitions to those rf-dressed states that have strong
inelastic processes when they collide with other atoms. As an atom moves fast
enough in the spatially inhomogeneous magnetic field, its spin $\vec{F}$ can
not follow adiabatically the changing magnetic field. The breakdown of the
adiabatic following is especially likely when the rf-field brings the bare
states into resonance. Any atom that has enough energy to leave the trap
during evaporative cooling will necessarily traverse such a resonance point.

We model the evaporation process by describing the atom as a wave packet. This
description takes into account any nonadiabatic process due to the atomic
motion as the wave packet moves back and forth in the trap. In other words, we
use the time-dependent Schr{\"o}dinger equation
\begin{equation}
   i\hbar\frac{\partial\Psi(\vec{x},t)}{\partial t} =
   {\cal H}(\vec{x})\Psi(\vec{x},t),\label{Erwin}
\end{equation}
where the state vector $\Psi(\vec{x},t)$ has $2F+1$ components (five in our
case), and ${\cal H}$ is the Hamiltonian consisting of the kinetic energy term,
the bare-atom Hamiltonian and the rf-induced couplings. Equation~(\ref{Erwin})
is then solved for some suitable initial condition $\Psi(\vec{x},0)$.

The true dynamics of evaporative cooling is quite complex, and we can
not realistically expect that Eq.~(\ref{Erwin}) describes the
situation completely, or that it can be easily solved when all the
characteristics of the situation are entered in to the atomic Hamiltonian
${\cal H}$.  We can, however, make simplifications, some of which have been
discussed in the previous section, that nevertheless allow us
to convincingly demonstrate some aspects of the rf-induced evaporative
cooling, including the nonadiabatic transitions.  We
restrict our studies to such times that the atoms can re-enter the
trap only once.  This ignores the possibility of multiple passes,
i.e., the possibility that those hot atoms which due to nonadiabatic
transitions return to the trap center can reach the rf-resonance again
at a later time.  For $^{23}$Na this approximation is allowed as the rate
coefficients for the inelastic processes suggest that only few such atoms
can survive until they reach the resonance region again.

Another simplification is that we consider only one spatial dimension, which
reduces the situation to the one shown in Fig.~\ref{pots}. However, this is
not a crucial simplification. In most cases during the evaporation process
the atomic wave packets behave very classically. In this picture only the
motion
of the atomic wave packet in the direction perpendicular to the rf-induced edge
of the trap is needed in studying the nonadiabatic transitions. In other words,
we need to model only that perpendicular motion. Although we allow our wave
packets to start from the center of the trap, at the edge of the trap their
behavior corresponds also to those atoms that have a more complicated 3D
motion but the same behavior in the perpendicular direction near the trap
edge. This also means that our approach is valid for all kinds of trap
configurations.

For $F=2$ we get in the bare state basis [from Eqs.~(\ref{sigma})
and~(\ref{coupl})]
\begin{equation}
   {\cal H}=\left( \begin{array}{ccccc}
   T -2\hbar\sigma(x) & \hbar \Omega & 0 & 0 & 0 \\ \hbar
   \Omega&T-\hbar\sigma(x)&\sqrt{\frac{3}{2}}\hbar \Omega &0&0\\
   0 & \sqrt{\frac{3}{2}}\hbar\Omega & T & \sqrt{\frac{3}{2}}\hbar\Omega & 0 \\
   0&0&\sqrt{\frac{3}{2}}\hbar\Omega&T+\hbar\sigma(x)&\hbar\Omega\\
   0&0&0&\hbar\Omega&T+2\hbar\sigma(x)
   \end{array}\right)
   \quad
   \begin{array}{l}
      |2,-2\rangle_z \\
      |2,-1\rangle_z \\
      |2,0\rangle_z \\
      |2,1\rangle_z \\
      |2,2\rangle_z
   \end{array};
   \qquad T=-\frac{\hbar^2}{2m}\frac{\partial^2}{\partial x^2}.
   \label{Hamilton}
\end{equation}
We take the mean momentum of the wave packet at the center of the trap as a
measure of the kinetic energy of the wave packet. In our study we have selected
the radiofrequency such that the crossing of the potentials occurs at the
energy corresponding to the kinetic energy of 5 $\mu$K in temperature units;
see Fig.~\ref{pots}. This sets our bare state trap depth $E_d/k_B=5\ \mu$K.
This is a rather low value, but it is difficult to increase it much more,
because the requirements for spatial and temporal step sizes in the numerical
treatment of Eq.~(\ref{Erwin}) increase rapidly with $E_d$. However, we shall
later discuss how our results can be scaled into other values of $E_d$ using
the multistate Landau-Zener model.

Our trap frequency is $112$ Hz, which sets the location of the crossing
point at 86 $\mu$m from the trap center. The initial wave packet is a Gaussian, 
which is reasonably narrow in both position and momentum, and moves towards the
crossing point. Since the initial kinetic
energy in our simulations is of the order of 5 $\mu$K, it is possible to
obtain narrow distributions in both representations. We distribute the Gaussian
wave packet over the bare states so that in fact only the lowest
rf-dressed state $a$ is occupied. Similarly, we express the final state
populations in terms of the rf-dressed states. However, the five-state problem
is easy to define in terms of the bare states, and thus we use this basis for
the numerical work. We have propagated the wave packets by splitting
the Hamiltonian~(\ref{Hamilton}) into a kinetic energy part and a potential
energy and coupling part (the split operator method). The kinetic energy part
is then treated using the Crank-Nicholson method, and the rest of $\cal H$ by
the diagonalization method. For more details see Ref.~\cite{Garraway95}.

\begin{figure}[thb]
\centerline{
\psfig{width=130mm,file=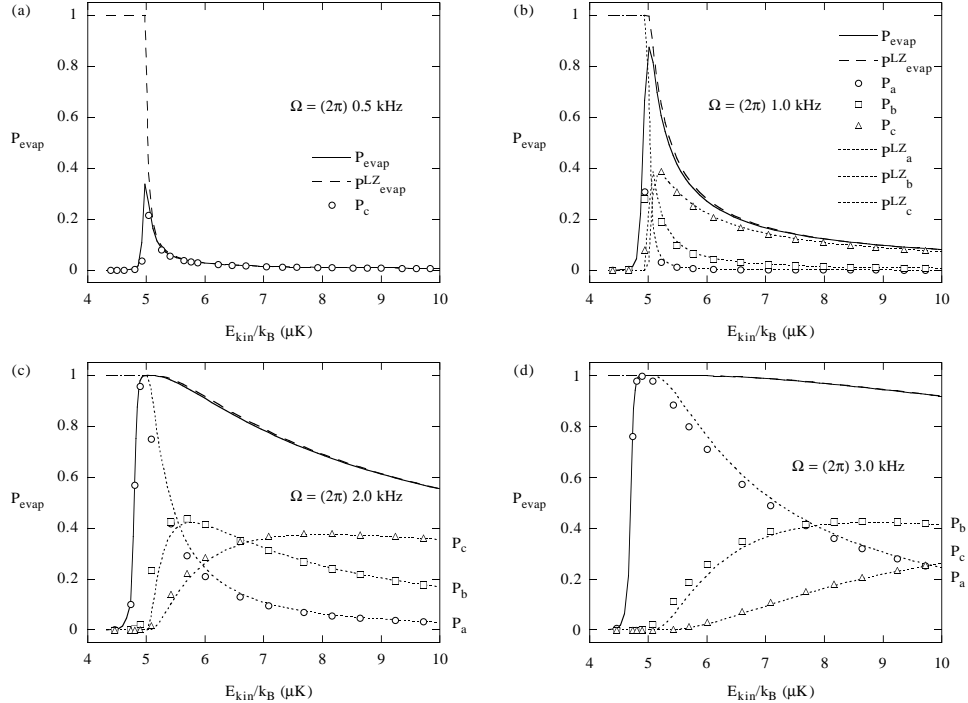}
}
\vspace{2mm}
\caption[f2]{The evaporation probability $P_{\rm evap}$ as a function of the
initial kinetic energy of the Gaussian wave packet for a selection
of Rabi frequencies after a single passage through the rf-induced resonance. We
show the wave packet results for total evaporation (solid line), and the
individual rf-dressed state contributions $P_a$, $P_b$ and $P_c$ when they are
nonnegligible. The dotted lines give the predictions by the multistate
Landau-Zener model for the probability of escaping the trap for the rf-dressed
states $a$, $b$ and $c$.  The dashed lines, labeled
$P^{\rm LZ}_{\rm evap}$, represent the total evaporation probability given by 
the multistate Landau-Zener model.  \label{evap}}
\end{figure}

In Fig.~\ref{evap} we demonstrate how the evaporation works as a function
of the kinetic energy $E_{\rm kin}$ of the initial wave packet for different
rf-coupling strengths $\Omega$ (as defined in Eq.~(\ref{coupl})). The curves
indicate the probability of the atoms leaving the trap after the first passage
of the crossing point by reaching positions $|x|\gg 86\ \mu$m. In addition to
showing the total evaporation $P_{\rm evap}$, we also give the contributions
($P_a$, $P_b$ and $P_c$) of the individual nontrapped (for
$|x|\gg 86\ \mu$m) rf-dressed states $a$, $b$ and $c$. Note that
$P_{\rm evap}=P_a+P_b+P_c$. For small $\Omega$ the
evaporation probabilities are clearly peaked at the resonance temperature
$E_d=5\ \mu$K. Also the flat state $c$ dominates the evaporation process,
demonstrating nonadiabaticity due to the insufficient coupling strength.
However, as $\Omega$ is increased, the leading contribution comes from the $a$
state. But although for e.g.~$\Omega=(2\pi)\ 3.0$ kHz the total evaporation is
nearly complete for a large range of temperatures, the contribution from $P_b$
and $P_c$ is still nonnegligible.

Typically one models the level crossings using the Landau-Zener theory, which
applies to two-state systems. However, here we have a multistate bowtie
crossing model, in which the states are coupled sequentially as in
Eq.~(\ref{coupl}). Fortunately, it has been shown that for the particular case
of Eqs.~(\ref{sigma}) and~(\ref{coupl}) the bowtie model can be solved
analytically~\cite{Kazansky96,Ostrovsky97,Vitanov97}. In fact, this solution
can be expressed in terms of the solutions to the related two-state
Landau-Zener (LZ) model, where
\begin{equation}
   H_{\rm LZ} = \frac{\hbar}{2}\left( \begin{array}{rr}
   -\sigma&\Omega\\ \Omega&\sigma \end{array}\right) \label{2LZ}
\end{equation}
According to the five-state case of the model (see e.g.~Ref.~\cite{Vitanov97})
the populations for the untrapped states after one traversal of the
crossing are
\begin{eqnarray}
   P_a &=& p^4,\nonumber\\ P_b &=& 4(1-p)p^3,\label{NLZ}\\ P_c &=&
   6(1-p)^2p^2,\nonumber
\end{eqnarray}
where $p= 1-\exp(-\pi\Lambda)$ is the probability for adiabatic
following in the two-state LZ model of Eq.~(\ref{2LZ}). The adiabaticity
parameter $\Lambda$ is defined for the two-state model as
\begin{equation}
   \Lambda = \frac{\hbar\Omega^2}{2\alpha v_C},\qquad \alpha=\hbar
   \left. \frac{d\sigma}{dR}\right|_{R=R_C}\label{Lambda},
\end{equation}
where $v_C$ is the speed of the atom at the rf-induced resonance point $R_C$.
It immediately follows that the Landau-Zener model is only applicable when
the total energy is higher than the bare-state energy at the resonance point.
For more details about applying LZ theory to wave packet dynamics see
Refs.~\cite{Suominen93,Suominen94,Garraway95}.

For the harmonic trap the atomic mass dependence cancels
in Eq.~(\ref{Lambda}), and we obtain
\begin{equation}
   \Lambda = 12.0\frac{1}{F}\frac{[\Omega(\mbox{kHz})]^2}
   {\sqrt{E(\mu\mbox{K})E_d(\mu\mbox{K})[\omega(\mbox{Hz})]^2}}\,,
   \label{Lambda2}
\end{equation}
where $E=E_{\rm kin}-E_d$ and $E_{\rm kin}$ is the kinetic energy of an atom
at the trap center. In Fig.~\ref{evap} we show the LZ
predictions as dotted lines together with the actual wave packet results. The
agreement is good for $E_{\rm kin}> 5 \mu$K, i.e., when $v_C$ is defined. For
smaller kinetic energies the Landau-Zener model cannot be applied, because the
model assumes a classical trajectory through the resonance, and for $E_{\rm
kin}<E_d$ this trajectory does not exist.

\begin{figure}[thb]
\centerline{
\psfig{width=86mm,file=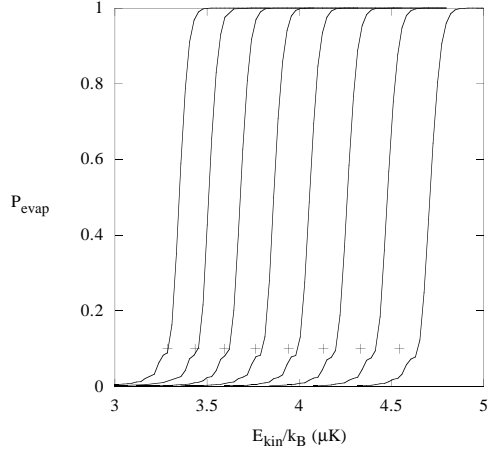}
}
\vspace{2mm}
\caption[f3]{The wave packet results for the total evaporation probability
$P_{\rm evap}$ as a function of the initial kinetic energy. The values of the
coupling $\Omega/(2\pi)$ are from right to
left: $2.5,5,7.5,10,12.5,15,17.5,20$ kHz. The plus-signs mark the
corresponding energy values where the barrier height for the rf-dressed states
is located in energy.  \label{evap-tanh}}
\end{figure}

In Fig.~\ref{evap-tanh} we show the evaporation probability as a 
function of the initial kinetic energy, expressed in temperature 
units, for values of $\Omega$ large enough that the evaporation is very 
adiabatic since nearly all atoms leave the trapping region in the 
rf-dressed state $a$.  The graphs all exhibit a rapid change from 
trapping into total evaporation around some critical kinetic energy.  
Since the increase in $\Omega$ leads to the lowering of the adiabatic 
barrier (see Fig.~\ref{pots}) and the atoms follow the rf-dressed 
states adiabatically, the critical temperature decreases with 
increasing $\Omega$.  However, if we consider the atomic motion in the 
lowest rf-dressed trapping state classically, we find that the minimum 
kinetic energy required for total evaporation is larger than the 
energy required to reach the top of the adiabatic barrier.  The top of 
these barriers is indicated by the plus-signs in Fig.~\ref{evap-tanh}.  
Note that eventually, as $E_{\rm kin}$ increases, the evaporation 
probability, although not shown here, will again drop.

\begin{figure}[thb]
\centerline{
\psfig{width=86mm,file=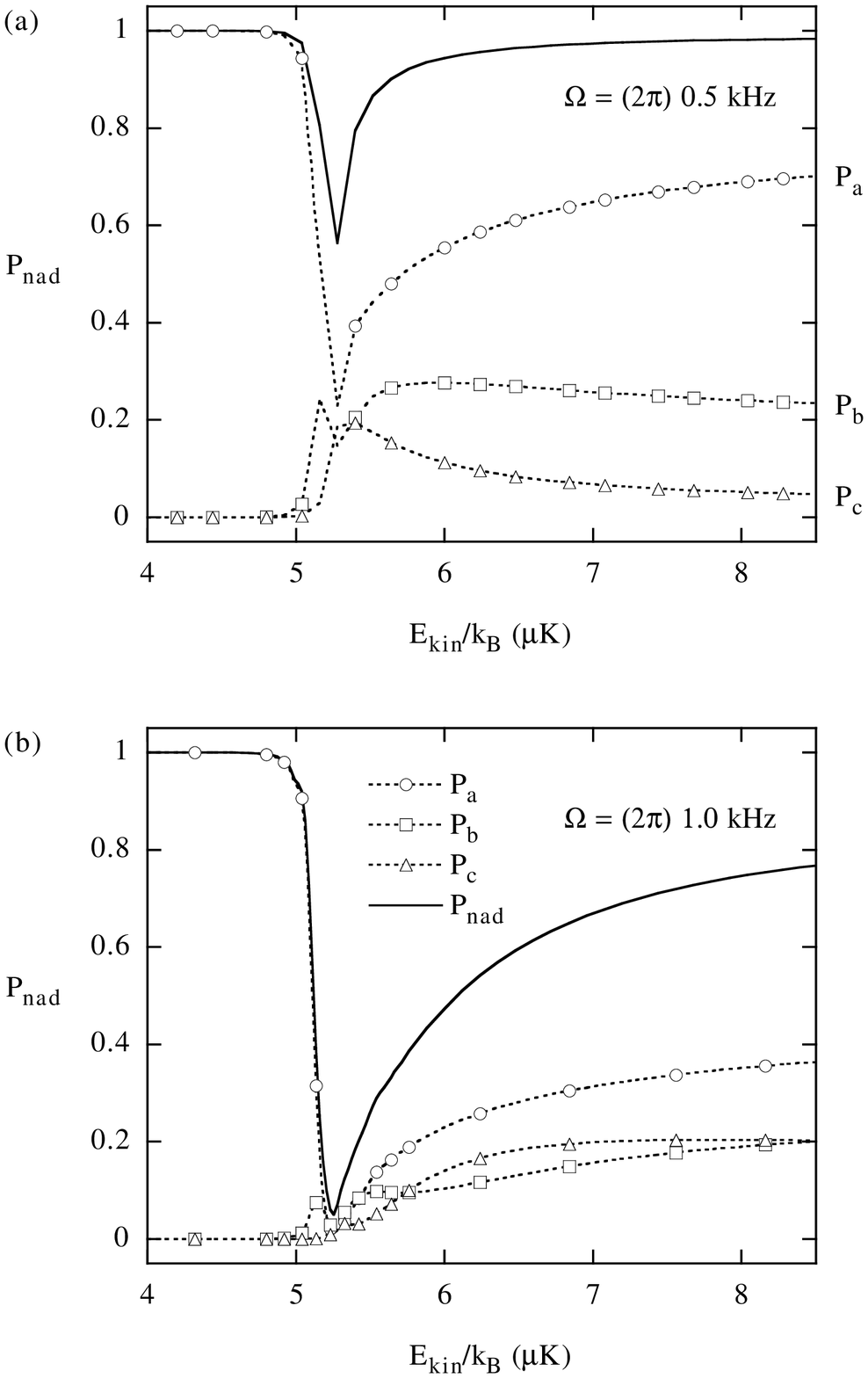}
}
\vspace{2mm}
\caption[f4]{The nonadiabatically induced population $P_{\rm nad}$ of the three
lowest rf-dressed states ($a$, $b$ and $c$) after a reflection from the
rf-resonance point as a function of the initial kinetic energy. The solid line
indicates the summed population of the three states. 
\label{nad}}
\end{figure}

For many values of the Rabi frequency and kinetic energy the evaporation is
not 100~\% efficient. Atoms can reflected back into the trap by
nonadiabatic transitions in which the atomic wave packet initially on 
state $a$ encounters the crossing region and is reflected back towards 
trap center in states $a$, $b$ or $c$.  The term nonadiabatic is now 
used in a slightly different fashion to imply that not only the spin 
state changes, as in the discussion of Fig.~\ref{evap}, but also the 
direction of the atom has changed.  Fig.~\ref{nad} illustrates for two 
different values of $\Omega$ the probability for returning to the trap 
center in states $a$, $b$, and $c$.  Clearly, if the field is not 
strong enough, a notable portion of the returning atoms can enter the 
unwanted $b$ and $c$ states, which correspond to the ($2,1$) and 
($2,0$) states at the trap center.  These nonadiabatic transitions 
where the atoms return to the trap center are not a direct problem if 
the collisional loss rates are small enough, since the trapped 
atoms can be allowed to "slosh" in the trap several times before they 
exit properly, as discussed by Ketterle and van 
Druten~\cite{Ketterle96}.  However, for $^{23}$Na $F=2$ atoms, these 
nonadiabatic transitions can cause complications, due to the large 
inelastic collision rates presented in Sec.~\ref{rates}.

Simple estimates can be made for the effect of inelastic collisions 
due to the hot returning $a=(2,1)$ and $b=(2,0)$ atoms inside the 
trap.  The characteristic time for collisions for rate constant $K$ 
and density $n$ is $\tau=1/(Kn)$ and the characteristic mean free path 
is $x_{\rm mfp} = 1/(\sigma n) = v/(Kn)$, where $\sigma$ is the collision 
cross section and $v$ the relative velocity.  With typical condensate 
densities on the order of $10^{14}$ atoms/cm$^{3}$, $\tau$ is on the 
order of a typical trap vibrational period and $x_{\rm mfp}$ on the order 
of the trap size for the case of $^{23}$Na, whereas $\tau$ is much 
longer than a vibrational period for the case of $^{87}$Rb.  
Therefore, a returning ($2,0$) $^{23}$Na atom is likely to undergo a 
``bad'' inelastic collision that ejects two atoms from the trap, 
whereas a ($2,0$) $^{87}$Rb atom would pass through the trap 
unhindered.  There could even be problems with collisions in a cold 
thermal gas above the critical temperature for Bose-Einstein 
condensation.  If nonadiabatic reflection produced ($2,1$) atoms, they 
would be trapped, and their density could build up.  Near the start of 
evaporation, where the atoms are still relatively hot, $p$-wave 
collisions with ($2,0$) atoms could result in atom loss, whereas near 
the end of evaporation where the collision energy is low enough that 
only $s$-wave collisions are significant, a build up of $(2,1)$ 
population would result in losses due to collisions between $(2,1)$ 
atoms with a time constant determined by the $(2,1)$ density.  This 
could easily be less than 1 s for $^{23}$Na even if only a small fraction 
of the $(2,2)$ population were converted to $(2,1)$.  These 
considerations suggest that evaporative cooling of $^{23}$Na in the 
$(2,2)$ state may be much more difficult than for $^{87}$Rb, for which 
evaporative cooling is known to work \cite{Anderson95}.  The rf power 
in the $^{23}$Na case would have to be sufficiently high to avoid 
nonadiabatic trap dynamics.

The multistate Landau-Zener model cannot describe the sequence of classical 
trajectories that leads to nonadiabatic reflection back into the center of 
the trap.  This setback is not very important, because it is clear that if 
the evaporation process is effective for a single pass, then necessarily we 
can ignore any nonadiabatic effects.  Since our wave packet 
results show that the multistate Landau-Zener theory is a good model 
for evaporation without reflection, we can always determine for any 
temperature and trap frequency the minimum rf-field strength that is 
required to quench the nonadiabatic processes by making the 
single-pass evaporation nearly 100 \% effective.  This is achieved by
requiring that $\Lambda\gg 1$ in Eq.~(\ref{Lambda2}).

\section{Conclusions}\label{conclusions}

We have shown that in the $^{23}$Na $F=2$ system inelastic
low-temperature collision rate coefficients can be larger than
$10^{-11}$ cm$^3$/s.  These rate coefficients occur when the
collision partners are (2,1) and (2,1), or (2,2) and (2,0).  For
$^{87}$Rb we find that all rate coefficients are small, on the order of
$10^{-14}$ cm$^3$/s.  Another interesting outcome of our studies is the
fact that rf-induced mixing plays no role in the occurrence of the
inelastic processes.  The rf-dressed states are obtained from the bare
states by a rotation.  However, this basis change does not alter the
physics of the situation, it merely redefines the quantization axis
for the atomic angular momentum.

In the absence of any mixing effects the inelastic processes in the $^{23}$Na
$F=2$ system can occur only if we have true occupation of the rf-dressed
states labelled as ($2,1$) or ($2,0$). Such occupation can be created by
nonadiabatic population transfer from the ($2,2$) rf-dressed state to the
other rf-dressed states. As the atoms move in spatially inhomogeneous fields,
the local orientation of the quantization axis ($B$-field) may change faster
than the ability of the system to follow it adiabatically. With wave packet
simulations we have demonstrated that the rf-field can induce such nonadiabatic
transfer during evaporative cooling. Furthermore, it is clear that due to this
transfer atoms can return to the trap center while occupying the rf-dressed
states ($2,1$) and ($2,0$).

The above effect can be reduced by keeping the strength of the rf-field so
large that the adiabatic following is ensured. However, at the early stages of
evaporative cooling the spatial change of potentials near the field-induced
resonance can be rapid. This can make it rather difficult to achieve the total
adiabatic following. In addition the $p$-wave contributions to the inelastic
processes are nonnegligible when $T\gtrsim 100\ \mu$K (for $^{23}$Na). Under
these circumstances one can not allow the hot atoms to slosh in the trap too
much before they leave it. So far condensation has not been demonstrated for
the $^{23}$Na $F=2$ system. We believe that, in order to achieve it, one has to
use clearly stronger rf-fields than e.g.~for the $^{23}$Na $F=1$ system.

Our simulations have also demonstrated that the multistate Landau-Zener
model~\cite{Kazansky96,Ostrovsky97,Vitanov97} can predict the basic wave packet
dynamics of the evaporation process correctly. This multistate model can be
given a simple analytic solution only when we have equal energy differences
between subsequent states and the sequential couplings given in
Eq.~(\ref{coupl}). If the rf-dressing of the states could not be described by
the simple rotation of the quantization axis, then the solution of the model
would have to include the complicated construction of the continuously changing
dressed states from the bare states, in addition to the actual nonadiabatic
processes. This is clearly connected to the fact that a simple solution for
the multistate LZ model in terms of the two-state case exists only for this
particular case. Despite its limitations, however, the multistate LZ model
gives us a simple and reliable criterion for making evaporation so efficient 
that nonadiabatic transitions are strongly quenched.

Recently it was shown experimentally that a $^{87}$Rb ($1,-1$) condensate can
be converted into a ($2,1$) condensate with a microwave pulse~\cite{Matthews98}.
Our results for $^{23}$Na ($2,1$)+($2,1$) collisions suggest that a similar
experiment with the $^{23}$Na system would not work. However, such an
experiment might be used to verify the loss rates presented in this paper, as
the pulsed creation of a ($2,1$) condensate allows one to set a distinctive
time point for the beginning of its destruction.

In addition to the process illustrated in Sec.~\ref{nonadiab} nonadiabatic
transfer can also arise if the frequency of the rf-field is not changed slowly
compared to the other time scales of the system. The moving atoms would then
see a chirped field, and the speed $v_C$ in the Landau-Zener model would
correspond more to the change of location of the rf-resonance point, than
to the
actual atomic wave packet motion through the resonance. This chirped rf-field
case could be studied with wave packet dynamics~\cite{Paloviita95}. It is also
used in the output coupler schemes for the trapped condensates~\cite{Mewes97}.
However, we have assumed that the rf-field frequency $\nu$ is changed so slowly
that this effect does not arise.

A similar situation, i.e., motion of the crossing point past the atoms, can
occur in the TOP trap, as discussed in Sec.~\ref{rates}~\cite{BillKris}. It is
generally assumed that the TOP field changes so fast that the passing of the
resonance is completely diabatic---this also means that the rf-field can be
ignored when the time-averaged field is determined for the TOP trap. If the
passing of the resonance were strongly adiabatic, the time-average would have
to be performed {\it after} adding the rf-induced resonances and couplings.
Then very interesting effects might arise (for an example see
Ref.~\cite{Dum98}).

\acknowledgments

This research has been supported by Army Research Office, Office for Naval
Research and the Academy of Finland.

\begin{references}
\bibitem{bare}  Here ``bare'' states are atom-field product states
                with energy equal to that for the atom and the field
                before taking into account the atom-field interaction, and
                ``dressed'' states are those which result from
diagonalizing the
                Hamiltonian including the atom-field interaction.
\end{references}
\end{document}